\documentclass[pre,showpacs,twocolumn]{revtex4}
\usepackage{graphicx}
\usepackage{units}

\begin{document}
\title{Quantum stress in chaotic billiards}
\author{Karl-Fredrik Berggren$^1$, Dmitrii N. Maksimov$^2$,
Almas F. Sadreev$^{1,2}$} \affiliation {1) IFM - Theory and Modeling, Link\"{o}ping University, S-581 83 Link\"{o}ping, Sweden\\ 2) L.V. Kirensky Institute of Physics, 660036, Krasnoyarsk, Russia}
\author{Ruven H\"{o}hmann, Ulrich Kuhl, Hans-J\"{u}rgen St\"{o}ckmann,}
\affiliation{AG Quantenchaos, Fachbereich Physik der Philipps-Universit\"{a}t Marburg,
Renthof 5, D-35032 Marburg, Germany}

\begin{abstract}
This article reports on a joint theoretical and experimental study
of the Pauli quantum-mechanical stress tensor $T_{\alpha \beta}(x,y)$
for open two-dimensional chaotic billiards. In the case of a finite
current flow through the system the interior wave function is expressed
as $\psi = u+iv$. With the assumption that $u$ and $v$ are Gaussian random fields
we derive analytic expressions for the statistical distributions for the
quantum stress tensor components $T_{\alpha \beta}$.
The Gaussian random field model is tested for a Sinai billiard with two
opposite leads by analyzing the scattering wave functions obtained numerically
from the corresponding Schr\"{o}dinger equation.
Two-dimensional quantum billiards may be emulated from planar microwave analogues.
Hence we report on microwave measurements for an open 2D cavity and how the
quantum stress tensor analogue is extracted from the recorded electric field.
The agreement with the theoretical predictions for the distributions for
$T_{\alpha \beta}(x,y)$ is quite satisfactory for small net currents.
However, a distinct difference between experiments and theory is observed
at higher net flow, which could be explained using a Gaussian random field,
where the net current was taken into account by an additional plane wave
with a preferential direction and amplitude.
\end{abstract}


\date{\today }
\maketitle

\section{Introduction}
\label{intro}

Chaotic quantum systems have been found to obey remarkable
universal laws related to, e.\,g.~, energy levels, eigenfunctions,
transition amplitudes or transport properties. These laws are independent
of the details of individual systems and depend only on spin and time-reversal
symmetries. The universality manifests itself in various statistical
distribution, such as the famous Wigner-Dyson distribution for the energy
levels in closed systems, the Thomas-Porter distribution for wave function
intensities, wave function form, conductance fluctuations, etc.\
(for overviews, see e.\,g.~[\onlinecite{stoe99,guh98,ber01a,fyo05b}]).
Two-dimensional ballistic systems like chaotic quantum billiards (quantum dots)
have played an important role in the development  of quantum chaos.
These systems are ideal because they have clear classical counterparts.
Nano-sized planar electron billiards may be fabricated from high-mobility
semiconductor hetero-structures like gated modulation-doped GaAs/AlGaAs and
external leads may be attached for the injection and collection of charge carriers
[\onlinecite{dav98}]. In this way one may proceed continuously from
completely closed systems to open ones. Here we will focus on open chaotic systems
in which a current flow is induced by external means. Simulations for open
chaotic 2D systems have shown, for example, that there is an abundance of chaotic
states that obey generalized wave function distributions that depend on
the degree of openness [\onlinecite{sai02b,kim05a}]. There are universal
distributions and correlation functions for nodal points and vortices
[\onlinecite{ber00b,sai01,ber02a,den07}] and the closely related universal
distributions [\onlinecite{sai02b,sad04}] and correlation functions for
the probability current density [\onlinecite{bar02,kim03b}].

In this article we will focus on the Pauli quantum stress tensor (QST)
for open planar chaotic billiards and its statistical properties. As we will see QST
supplements previous studies of wave function statistics and flow patterns in an
important way as it probes higher order derivatives (irrespective of
the chosen gauge) and thereby fine details of a wave function. QST was
introduced by Pauli [\onlinecite{pau33,pau80}] already in 1933 but in contrast
to the corresponding classical entities for electromagnetic fields and fluids
[\onlinecite{mis73}], for example, it has remained somewhat esoteric since then.
On the other hand, studies of stress are in general an important part of material
science research and, on a more fundamental atomistic level, stress originates from
quantum mechanics. Efficient computational methods based on electronic structure
calculations of solids have therefore been developed to analyze both kinetic and
configurational contributions to stress [\onlinecite{nie85,fol86,god88}]. The
recent advances in nanomechanics also puts more emphasis on the quantum-mechanical
nature of stress [\onlinecite{she04}]. Furthermore it features in quantum hydrodynamic
simulations of transport properties of different quantum-sized semiconductor devices
like Resonant Tunneling Devices (RTD) and High Electron Mobility Transistors (HEMT)
[\onlinecite{hoen04}] and in atomic physics and chemistry [\onlinecite{god90,tao1}].
All in all, QST is a fundamental concept in quantum mechanics that ties up with local forces
and the flow of probability density. Hence it is natural to extend the previous studies
of generic statistical distributions for open chaotic quantum billiards to also include
the case of stress. Our choice of planar ballistic quantum billiards is favorable in this
respect as stress is then only of kinetic origin. Moreover, the motion in an open
high-mobility billiard may ideally be viewed as interaction-free because the
nominal two-dimensional
mean free path may exceed the dimensions of the
billiard itself. In this sense we are dealing to a good approximation with single-particle behavior.

There is an ambiguity in the expression for the stress tensor because any
divergence-free tensor may be added without affecting the forces [\onlinecite{rog02,mar07}].
For clarifying our definitions and particular choice, we repeat the basic steps,
albeit elementary, in Pauli's original derivation of his QST [\onlinecite{pau33,pau80}].
If $\psi(\bf{x},t)$ is a solution to the Sch\"odinger equation
\begin{equation}
 i\hbar \frac{\partial \psi}{\partial t}=-\frac{\hbar^2}{2m}\Delta \psi +V\psi,
 \label{schrodinger}
\end{equation}
for a particle with mass $m$ moving in the external potential $V$, the components of the
probability current density are
\begin{equation}
 j_{\alpha}=\frac{\hbar}{2mi}\left(\psi^{*}\frac{\partial \psi}{\partial x_{\alpha}}
 -\psi\frac{\partial \psi^{*} }{\partial x_{\alpha}}\right).
 \label{curr}
\end{equation}
Taking the time derivative of $j_{\alpha}$ and using the the right hand side of the
Schr\"odinger equation above to substitute $\partial \psi/\partial t$, Pauli arrived at
the expression
\begin{equation}
 m\frac{\partial {j_{\alpha}}}
 {\partial t}=-\sum_{\beta}\frac{\partial T_{\alpha
 \beta}}{\partial x_{\beta}}-\frac{\partial V}{\partial
 x_{\alpha}}|\psi|^2,
 \label{currentder1}
\end{equation}
where  $T_{\alpha \beta}$ is his form of the quantum-mechanical stress tensor
\begin{eqnarray}\label{2}
T_{\alpha \beta}=\frac{\hbar^2}{4m}
&&\left[ -\psi^{*}\frac{\partial^{2} \psi }{\partial x_{\alpha}
\partial x_{\beta} } - \psi\frac{\partial^{2} \psi^{*} }{\partial
x_{\alpha} \partial x_{\beta} } \right. \nonumber \\
&& + \left. \frac{\partial\psi}{\partial \
x_{\alpha}}\frac{\partial \psi^{*} }{\partial x_{\beta}} +
\frac{\partial\psi^{*}}{\partial \ x_{\alpha}}\frac{\partial \psi
}{\partial x_{\beta}} \right].
\end{eqnarray}
In case of planar billiards $V$ may be put equal to zero and it is in that form that
we will explore Eq.~(\ref{2}). The kinetic Pauli QST is sometimes referred to as the
quantum-mechanical momentum flux density, see e.\,g. Ref.~[\onlinecite{god88}].
>From now on we will simply refer to it as QST.

There are obvious measurement problems associated with QST
for a quantum billiard, among them the limited spatial resolution
presently available (see e.\,g.~Ref.~[\onlinecite{cro03}]). In
the case of 2D quantum billiards there is, however, a beautiful
way out of this dilemma, a way that we will follow here. It turns
out that single-particle states $\psi$ in a hard-wall quantum billiard with
constant inner potential obey the same stationary Helmholtz
equation and same boundary condition as states in a flat microwave
resonator do [\onlinecite{stoe99}]. This means that our quantum
billiard can be emulated from microwave analogues in which the
perpendicular electric field $E_z$ takes the role of the wave
function $\psi$. Since the electric field may be measured this
kind of emulation gives us a unique opportunity to inspect the
interior of a quantum billiard experimentally
\cite{sri91,ste92,lau94b,gok98,dem99,lau07}.
Using the one-to-one correspondence between the Poynting vector and the probability
current density, probability densities and currents have been
studied in a microwave billiard with a ferrite insert as well as
in open billiards. Distribution functions based on measurements
were obtained for probability densities, currents, and
vorticities. In addition, vortex pair correlation functions
have be extracted. For all quantities studied
[\onlinecite{fyo05b,bar02,kim03b}] complete agreement was obtained
with predictions based on the assumption that wave functions in an
chaotic billiard may be represented by a random superposition of
monochromatic plane waves [\onlinecite{ber77a}].

The layout of the article is the following. In Section~\ref{sectionII}
we outline the meaning of QST by referring to Madelung's hydrodynamic
formulation of quantum mechanics from 1927 [\onlinecite{mad27}].
Section~\ref{sectionIII} presents the derivation of the distribution functions
for the components of the QST in 2D assuming that the wave function may be
described in terms of a random Gaussian field and that the net current is zero.
Although our focus is on 2D the results are extended to 3D as well.
Section~\ref{sectionIV} deals with the distribution of the quantum potential that
appears naturally in the hydrodynamic formulation of quantum mechanics.
In Section~\ref{sectionV} we present numerical simulations of transport through
an open Sinai billiard with two opposite leads and a comparison with the analytical
Gaussian random field model is made. Microwave measurements are reported in
Section~\ref{exp} and analyzed in terms of the quantum stress tensor. A Berry-type
wave function with directional properties is introduced in the same section to analyze
the influence of net currents on the statistical distributions for $T_{\alpha \beta}(x,y)$.

\section{The meaning of QST}
\label{sectionII}

One of the earliest physical interpretations of the Schr\"{o}dinger
equation is due to Madelung who introduced the hydrodynamic
formulation of quantum mechanics already in 1927 [\onlinecite{mad27}].
This is a helpful step to get a more intuitive understanding in
classical terms of, for example, quantum-mechanical probability
densities and the meaning of quantum stress (see e.\,g.~Refs.
[\onlinecite{bia92,hol93,wya05}]). Madelung obtained the QM
hydrodynamic formulation by rewriting the wave function $\psi$ in
polar form as
\begin{equation}\label{polar}
\psi({\bf x},t)=R({\bf x},t)\, e^{iS({\bf x},t)/\hbar}.
\end{equation}
The probability density is then $\rho=R^2$. By introducing the
velocity ${\bf v}={\bf{\nabla}} S({\bf x},t)/m$ the probability
density current or probability flow is simply ${\bf j}\, = \,\rho{\bf v}$.
Intuitively this is quite appealing. Inserting the polar form in the
Pauli expression for $T_{\alpha \beta}$ in Eq.~(\ref{2}) we then have
\begin{equation}\label{polarstress2}
T_{\alpha,\beta}=\frac{\hbar^2}{4m}
\left(-\frac{\partial^2 \rho}{\partial x_{\alpha}x_{\beta}}
+\frac{1}{\rho}\frac{\partial \rho}{\partial x_{\alpha}}
\frac{\partial \rho}{\partial x_{\beta}}\right)
+\rho mv_{\alpha}v_{\beta}.
\end{equation}
There are two qualitatively different terms in Eq.~(\ref{polarstress2}),
a quantum-mechanical term $\tilde{T}_{\alpha\beta}$ that contains the
factor $\hbar$ and therefore vanishes in the classical limit $\hbar\rightarrow 0$,
plus the "classical" contribution $\rho mv_{\alpha}v_{\beta}$ which remains
in the classical limit. Using the notations above Eq.~(\ref{currentder1}) gives
the quantum hydrodynamic analogue of the familiar classical Navier-Stokes equation for the
flow of momentum density $m\rho{\bf v}$
\begin{equation}\label{momflow}
  m\frac{\partial\rho v_{\alpha}}
 {\partial t}=-\sum_{\beta} \nabla_{\beta} T_{\alpha \beta} - \rho
 \nabla_{\alpha} V.
\end{equation}
Alternatively the Schr\"odinger equation may be rewritten as the two familiar
hydrodynamic equations in the Euler frame [\onlinecite{bia92,hol93,wya05}]
\begin{eqnarray}
\frac{\partial\rho}{\partial t}+\nabla \cdot [\rho {\bf
v}]=0\,\,\\ \ \frac{\partial{\bf v}}{\partial t}+[{\bf
v}\cdot\nabla]{\bf v}={\bf f}/m+{\bf F}/m,
\end{eqnarray}\label{NSA}
where the external force is due to external potential
\begin{equation}\label{forceextA}
{\bf f}=-\nabla V,
\end{equation}
and the internal force is due to the quantum potential
\begin{equation}\label{forceintA}
{\bf F}=-\nabla V_{QM}, \qquad V_{QM}=-\frac{\hbar^2}{2m}\frac{\nabla^2
R}{R}.
\end{equation}
Then the internal force can be expressed by a stress tensor for
the probability fluid as
\begin{equation}\label{quantumstress}
F_{\alpha}=-\sum_{\beta}\frac{1}{\rho}\frac{\partial
\tilde{T}_{\alpha \beta}}{\partial x_{\beta}}.
\end{equation}

Thus we are dealing with a ``probability fluid" in which flowlines
and vorticity patterns are closely related to QST.

\section{Distribution of QST for a quantum billiard}
\label{sectionIII}

We now return to the full expression for the stress tensor $T_{\alpha \beta}$
in Eq.~(\ref{2}). Consider a flat two-dimensional ballistic cavity (quantum dot)
with hard walls. Within the cavity we therefore have $V=0$ and the corresponding
Schr\"{o}dinger equation is $(\triangle+k^2)\psi(x,y)=0$ with $k^2=2mE/\hbar^2$,
where $k$ is the wave number at energy $E$. In this case the wave function may be
chosen to be real if the system is closed and, as a consequence, there is no
interior probability density flow. The wave function normalizes to one over the
area $A$ of the cavity. On the other hand, if the system is open, for example by
attaching external leads, and there is a net transport, the wave function must be
chosen complex. Thus
\begin{equation}\label{1}
 \psi\rightarrow u+iv,
\end{equation}
in which $u$ and $v$ independently obey the stationary Schr\"{o}dinger equation for
the open system. In the following discussion it is convenient to make a substitution
to dimensionless variables, ${k\bf x}\rightarrow {\bf x}'$. Hence we have
$(\triangle'+1)\,u(x',y')=0$ and similarly for $\,v(x',y')$. The size of the cavity
scales accordingly as $A\rightarrow A'$.

If the shape of the cavity is chaotic we may assume that $u$ and $v$ are to a good
approximation random Gaussian functions (RGFs) [\onlinecite{sai02b,mcd88}] with
$\langle u^2 +v^2 \rangle=1+\epsilon ^2$,
$\langle v^2\rangle=\epsilon ^2\langle u^2 \rangle$,
$\langle uv\rangle=0$ and $ ~\langle u\rangle= \langle v\rangle=0$.
If $u$ and $v$ were correlated we can apply a phase transformation [\onlinecite{sai02b}]
which makes these functions uncorrelated. Here we use the definition
\begin{equation}\label{norm}
\langle \ldots\rangle=\frac{1}{A}\int\ldots
dA=\frac{1}{A'}\int\ldots dA'.
\end{equation}
In what follows we thus use dimensionless derivatives in ${\bf x}'$ and express the
QST components in units of the energy $\hbar^2 k^2/2m$. From Eq.~(\ref{2}),
dropping the prime $ (')$ in the expressions from now on, we then have
\begin{equation}\label{3}
 T_{xx}=-u\frac{\partial^{2}u}{\partial^{2}x}
 -v\frac{\partial^{2}v}{\partial^{2}x}+{\left(\frac{\partial u}{\partial
 x}\right)}^2+{\left(\frac{\partial v}{\partial x}\right)}^2
\end{equation}
and
\begin{equation}\label{13}
 T_{xy}=-u\frac{\partial^{2}u}{\partial x \partial y}
 -v\frac{\partial^{2}v}{\partial x \partial y}+\frac{\partial u}{\partial
 x}\frac{\partial u}{\partial y} + \frac{\partial v}{\partial
 x}\frac{\partial v}{\partial y}.
\end{equation}

{\it Two-dimensional case.} Let us first consider the distributions of the stress tensor
for a two-dimensional complex RGF $\psi$. In the following derivation we assume that the
net current from one lead to the other is so small that in practice we are dealing
with isotropic RGFs. We therefore have
\begin{eqnarray}\label{4}
&&\langle uu_{xx} \rangle =-\frac{1}{2}, \,\,
\langle u_{x}^{2}\rangle =\frac{1}{2}, \,\,
\langle uu_{x} \rangle =0, \,\,\nonumber \\
&&\langle u_{x}u_{xx} \rangle =0, \,\,
\langle u_{xx}^{2} \rangle
=\frac{3}{8}
\end{eqnarray}
for the two-dimensional case.  The corresponding expressions for $v$ follow simply by
replacing $u, u_x, u_{xx}$ etc.  by $v/\epsilon, v_x/\epsilon, v_{xx}/\epsilon $ and
so on.

For the component $T_{xx}$ in Eq.~(\ref{3}) we need the following joint distribution
of two RGFs [\onlinecite{ebe84}]
\begin{equation}\label{5}
 f(\overrightarrow{X})=\frac{1}{2\pi
 \sqrt{\det(K)}}\exp\left[-\frac{1}{2}\overrightarrow{X}^{\dag}K^{-1}
 \overrightarrow{X}\right].
\end{equation}
where $\overrightarrow{X}^{\dag}=(u,v,u_x,v_x, u_{xx}, v_{xx})$, and the matrix
$K=\langle \overrightarrow{X}\overrightarrow{X}^{\dag}\rangle$. For an isotropic
RGF there are only correlations $\langle u u_{xx} \rangle,~\langle v v_{xx} \rangle$.
Therefore the only nontrivial block of the total matrix $K$ is the matrix
\begin{equation}\label{6}
K_u=\left(\matrix{1 & -1/2\cr -1/2 & 3/8\cr}\right), ~
K_u^{-1}=\left(\matrix{3 & 4\cr 4 & 8\cr}\right)
\end{equation}
for the RGFs $u, u_{xx}$ and the matrix $K_v=\epsilon K_u$ for the two RGFs for $v$
and $v_{xx}$. Correspondingly we obtain from Eq.~(\ref{5})
\begin{equation}\label{7}
 f(u,u_{xx})=\frac{\sqrt{8}}{2\pi} \exp\left\{
 -\frac{3u^{2}+8uu_{xx}+8u_{xx}^2}{2}\right\}
\end{equation}
and
\begin{equation}\label{fvvxx}
 f(v,v_{xx})=\frac{\sqrt{8}}{2\pi\epsilon^2} \exp\left\{-
 \frac{3v^{2}+8vv_{xx}+8v_{xx}^2}{2\epsilon^2}\right\}.
\end{equation}
The characteristic function of the stress tensor component $T_{xx}$ is
\begin{equation}\label{8}
 \theta(a)=\langle e^{iaT_{xx}} \rangle
\end{equation}
and takes the following explicit form
\begin{eqnarray}\label{9}
 \theta(a)&=&8[(1-ia)(1-i\epsilon a)
 \nonumber\\
 && (a-i(\sqrt{24}+4))(\epsilon a -i(\sqrt{24}+4))
 \nonumber\\
 &&  (a+i(\sqrt{24}-4))(\epsilon a
 -i(\sqrt{24}-4))]^{-1/2}.
\end{eqnarray}
As a result we obtain for the distribution function
\begin{equation}\label{10}
 P(T_{xx})=\frac{1}{2\pi}\int_{-\infty}^{\infty}\theta(a)e^{-iaT_{xx}} da.
\end{equation}

\begin{figure}[ht]
\includegraphics[width=.8\columnwidth]{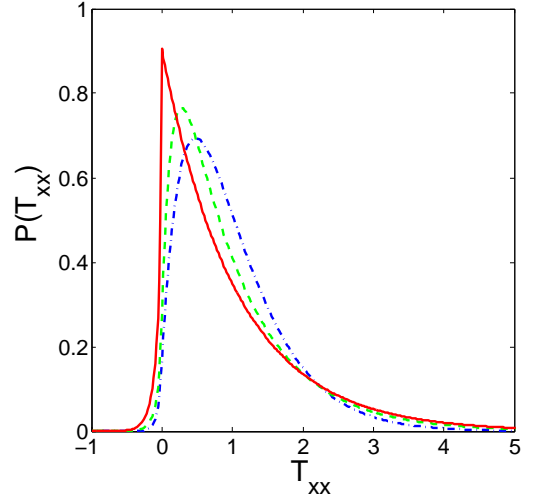}
\caption{ (color online)
The distribution $P(T_{xx})$ for $\epsilon=1$ (dash-dotted line), $\epsilon=0.5$
(dashed line), and $\epsilon=0$ (solid line). The stress tensor component $T_{xx}$
is measured in terms of the mean value $\langle T_{xx}\rangle$.}
\label{fig1}
\end{figure}

For $\epsilon\neq 1$ this integral may be calculated numerically. However for
$\epsilon=1$ it might be evaluated analytically. In particular, for $T_{xx}>0$
we obtain
\begin{equation}\label{11}
 P(T_{xx})=\frac{2}{\sqrt{6}} \frac{e^{-(\sqrt{24}-4)T_{xx}}}
 {(5-\sqrt{24})} - 8e^{-T_{xx}},
\end{equation}
and for $T_{xx}< 0$
\begin{equation}\label{12}
 P(T_{xx})=\frac{2}{\sqrt{6}} \frac{e^{(\sqrt{24}+4)T_{xx}}}
 {(5+\sqrt{24})}.
\end{equation}
The distribution~(\ref{12}) is shown in Fig.~\ref{fig1} together with results
for different $\epsilon$-values obtained by numerical evaluation of the
integral~(\ref{10}). Note that the distributions are here given in terms of
$\langle T_{xx}\rangle=1+\epsilon^2$.

To repeat the calculations for the component $T_{xy}$ we need the following
correlators
\begin{equation}\label{14}
\langle uu_{xy} \rangle =0, ~ \langle u_{x}u_{xy} \rangle =0,
\langle u_{y}u_{xy} \rangle =0, \langle u_{xy}^{2} \rangle
=\frac{1}{8}
\end{equation}
for the 2D case. The correlation matrix turns out to be diagonal. Then the
characteristic function
\begin{eqnarray}\label{15}
\theta(a)&=&2[(2+(a/2)^2)(2+(\epsilon a/2)^2)(1+(a/2)^2) \nonumber \\
&&(1+(\epsilon a/2)^2)]^{-1/2}
\end{eqnarray}
defines the distribution $P(T_{xy})$. For $\epsilon=1$ the integral~(\ref{10})
may, as above, be performed analytically to give
\begin{equation} \label{16}
 P(T_{xy})=2e^{-2|T_{xy}|}-\sqrt{2}e^{-2\sqrt{2}|T_{xy}|}.
\end{equation}

\begin{figure}[ht]
\includegraphics[width=.8\columnwidth]{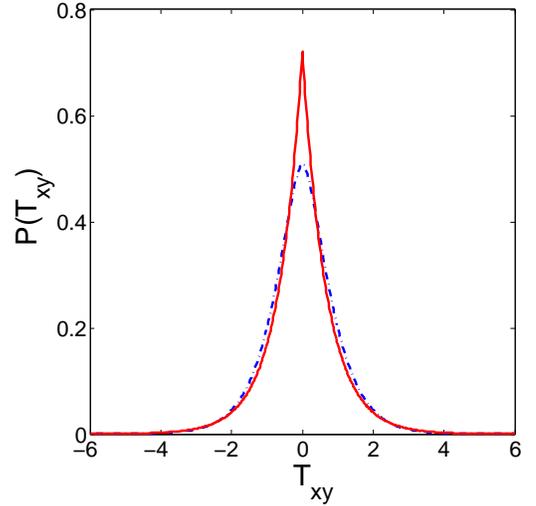}
\caption{ (color online)
The distribution $P(T_{xy})$ for $\epsilon=1$ (dash-dotted line) and
$\epsilon=0$ (solid line). The stress tensor component $T_{xy}$ is
measured in terms of mean value $\sqrt{\langle T_{xy}^2\rangle}$. }
\label{fig2}
\end{figure}

The distributions $P(T_{xy})$ in (\ref{16}) are shown in Fig.~\ref{fig2}
for the two cases $\epsilon=0$ and $\epsilon=1$. Only two cases are shown
because of the small differences in $P(T_{xy})$ for different $\epsilon$-values.
The distributions  are in this case given in terms of
$\sqrt{\langle T_{xy}^2\rangle}$, where $\langle T_{xy}^2\rangle=\frac{3}{8}(1+\epsilon^4)$.

{\it Three-dimensional case.}
In this case the expressions in~(\ref{4}) are to be replaced by
\begin{eqnarray}\label{3d}
&&\langle uu_{xx} \rangle =-\frac{1}{3}, \,\,
\langle u_{x}^{2} \rangle =\frac{1}{3}, \,\, \nonumber \\
&&\langle uu_{x} \rangle =0, \,\,
\langle u_{x}u_{xx} \rangle =0,\,\,
\langle u_{xx}^{2} \rangle
=\frac{1}{5}
\end{eqnarray}
and (\ref{14}) by
\begin{equation}\label{143d}
\langle uu_{xy} \rangle =0, ~ \langle u_{x}u_{xy} \rangle =0,
\langle u_{y}u_{xy} \rangle =0, \langle u_{xy}^{2} \rangle
=\frac{1}{15}.
\end{equation}
Accordingly the correlation matrix~(\ref{6}) is
\begin{equation}\label{Ku3d}
K_u=\left(\matrix{1 & -1/3\cr -1/3 & 1/5\cr}\right), \,\,
K_u^{-1}=\frac{1}{4}\left(\matrix{9 & 15\cr 15 & 45\cr}\right).
\end{equation}
The joint probability function of two RGFs $u$ and $u_{xx}$ then takes the
following form
\begin{equation}\label{fuuxx}
 f(u,u_{xx})=\frac{\sqrt{45}}{2\pi} \exp\left\{
 -\frac{9u^{2}+30uu_{xx}+45u_{xx}^2}{8}\right\}.
\end{equation}
The characteristic function defining the distribution $P(T_{xx})$ is
\begin{eqnarray}\label{thetaxx3d}
 \theta(a)=\frac{45}{(3/2-ia)(ia+15/4+9\sqrt{5}/4)} \nonumber \\
 \frac{1}{(ia+15/4-9\sqrt{5}/4)}
\end{eqnarray}
and, correspondingly,
\begin{equation}\label{FTxx3dr}
 P(T_{xx})=\frac{5}{(7\sqrt{5}-15)}e^{-\frac{9\sqrt{5}-15}{4}T_{xx}}
 - \frac{15}{2}e^{-\frac{3}{2}T_{xx}}
\end{equation}
for $T_{xx}>0$, and
\begin{equation}\label{FTxx3d}
P(T_{xx})=\frac{5}{(7\sqrt{5}+15)}e^{\frac{9\sqrt{5}+15}{4}T_{xx}}
\end{equation}
for $T_{xx}< 0$. Identical expressions hold for the two other diagonal
components.

In a similar way we obtain the distribution function for the off-diagonal
components $\alpha\neq \beta$. For the specific case $\epsilon=1$ we have
according to Eq.~(\ref{143d})
\begin{equation}\label{charxy3d}
\theta(a)=\frac{2}{[3+(a/2)^2][1+(a/2)^2]},
\end{equation}
and
\begin{equation}\label{Fxy3d}
 P(T_{xy})=\frac{15}{4}e^{-3|T_{xy}|}-\frac{3\sqrt{15}}{4}e^{-\sqrt{15}|T_{xy}|}.
\end{equation}
The expression for the other off-diagonal components are, of course,
identical.

\section{Distribution of quantum potential}
\label{sectionIV}

The quantum or internal force in Eq.~(\ref{forceintA}) in the hydrodynamic
formulation is defined by the quantum potential $V_{QM}$. In terms of the
RGFs $u, ~v$ it may be written as
\begin{equation}\label{VQMold}
 V_{QM}=-V_x-V_y,
\end{equation}
$$  V_x=\frac{uu_{xx}+vv_{xx}+u_x^2+v_x^2}{u^2+v^2}-
 \left(\frac{uu_x+vv_x}{u^2+v^2}\right)^2, $$
$$V_y=\frac{uu_{yy}+vv_{yy}+u_y^2+v_y^2}{u^2+v^2}-
\left(\frac{uu_y+vv_y}{u^2+v^2}\right)^2.$$
The second derivatives might be eliminated using the Schr\"odinger equations
for $u$ and $v$, i.\,e., $u_{xx}+u_{yy}=-u, ~~v_{xx}+v_{yy}=-v$. As a result
we have
\begin{equation}\label{VQM}
 V_{QM}=1-\frac{(uv_x-vu_x)^2+(uv_y-vu_y)^2}{{\rho}^2}.
\end{equation}
which implies
\begin{equation}\label{boundary}
-\infty \leq V_{QM}\leq 1.
\end{equation}
The distribution of the quantum potential is given by
\begin{equation}\label{FV}
 P(V_{QM})=\frac{1}{2\pi}\int \exp(-iaV_{QM})\theta(a)da,
\end{equation}
where
\begin{equation}\label{charV}
\theta(a)=\langle\exp(iaV_{QM})\rangle=\int
d^{6}\overrightarrow{X}f(\overrightarrow{X})\exp(iaV_{QM}),
\end{equation}
$f(\overrightarrow{X})$ is given by the same formula as~(\ref{5}), however,
with vector $\overrightarrow{X}^{+}=(u,v,u_x,v_x,u_y,v_y)$ with the same
correlators as~(\ref{6}).

For~(\ref{charV}) we may now write with $\epsilon=1$, which is the only case
accessible in closed analytic form,
\begin{equation}\label{charV1}
\theta(a)=\frac{1}{2\pi}\int dudv
\Gamma_x\Gamma_y\exp\left[-\frac{1}{2}(u^2+v^2)+ia\right],
\end{equation}
with
\begin{equation}\label{Gx}
 \Gamma_x=\frac{1}{\pi}\int du_x dv_x\exp\left\{
 -u_x^{2}-v_x^2+\frac{ia(uu_x-vv_x)^2}{\rho^2}\right\}.
\end{equation}
The same expression holds for $\Gamma_y$. The integration in~(\ref{Gx}) gives
\begin{equation}\label{GxGy}
 \Gamma_x\Gamma_y =\frac{-i\rho}{a-i\rho}.
\end{equation}
Substituting (\ref{GxGy}) into (\ref{charV1}) we obtain
\begin{equation}\label{charV2}
 \theta(a)=-i\int_{0}^{\infty}\frac{drr^3}{a-ir^2}\exp(ia-r^2/2)
\end{equation}
where $r=\sqrt{\rho}$. Finally, substituting that into (\ref{FV})
we obtain the distribution function for the quantum potential
\begin{equation}\label{FVfinal}
 P(V_{QM})=\frac{1}{2(3/2-V_{QM})^2}.
\end{equation}

\begin{figure}
\includegraphics[width=.8\columnwidth]{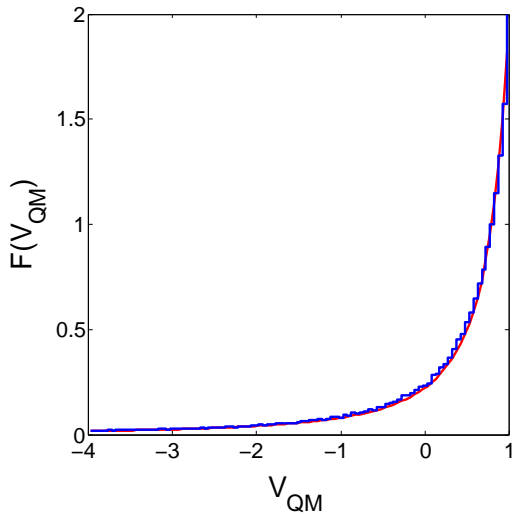}
\caption{(color online)
The distribution of the quantum potential (\ref{FVfinal}) for $\epsilon=1$
compared to numerical histogram based on the Berry function in Eq.~(\ref{Berry}).}
\label{fig3}
\end{figure}

The distribution (\ref{FVfinal}) is normalized as $\int_{-\infty}^1P(V)dV=1$.
The distribution of $P(V_{QM})$ is shown in Fig.~\ref{fig3} and compared to a
numerical computation of the same statistics based on the Berry conjecture for
chaotic wave functions [\onlinecite{ber77a}]
\begin{equation}\label{rw}
 \psi(r)=\frac{1}{\sqrt{A}}\sum_{n} a_n e^{i{\bf k}_n{\bf \cdot r}}.
 \label{Berry}
\end{equation}
Here $A$ is the area of the random monochromatic plane wave field with
$|{\bf k}_n|^2=1$ and the amplitudes for the random plane waves obey the relation
$\langle a_n^2 \rangle = \frac{1}{N}$. The Berry function in Eq.~(\ref{Berry})
corresponds to $\epsilon=1$.

\section{Numerical simulations of scattering states in an open chaotic electron billiard}
\label{sectionV}

A billiard becomes an open one when it is connected to external reservoirs,
for example, via attached leads. A stationary current through the system may
be induced by applying suitable voltages to the reservoirs (or by a microwave
power source as in Section~\ref{exp}). Here we consider hard-walled Sinai-type
billiards with two opposite normal leads. A first step towards a numerical
simulations of the quantum stress tensor is to find the corresponding scattering
states by solving the Schr\"{o}dinger equation $-\nabla^2 \psi= k^2\psi$ for the
entire system. The numerical procedure for this is well known. Thus we use the
finite difference method for the interior of the billiard in combination with
the Ando boundary condition [\onlinecite{ber02a,and91}] for incoming, reflected
and transmitted solutions in the straight leads. Once a scattering wave function
has been computed in this way the fraction  residing in the cavity itself is
extracted for the statistical analysis. To ensure statistical independence of the
real and imaginary parts $u$ and $v$ a global phase is removed as discussed in
Ref.~[\onlinecite{sai02b}]. By this step we also find the value of $\epsilon$.
The interior wave function is then normalized as defined in Section~\ref{sectionIII}.

\begin{figure}[ht]
\includegraphics[width=\columnwidth]{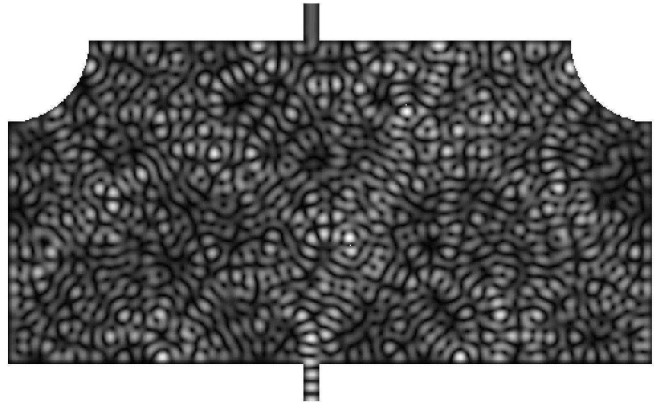}
\caption{
View of the scattering wave function in the open Sinai billiard for the case (A) shown
in Fig.~\ref{transm} for $k^2=30.878$ in dimensionless units (see text) and for small
aspect ratio $d/L$ = 2/67 (ratio between the widths of the leads and the billiard).
The system is asymmetric because the two opposite leads are slightly off the middle
symmetry line of the nominal billiard. Only the lowest channel is open in the leads.}
\label{psi}
\end{figure}

For the numerical work it is convenient to make the substitution $x\rightarrow x/d$
and $y\rightarrow y/d$ where $d$ is the width of the leads. Here we use dimensionless
energy $k^2=E/E_0, E_0=\hbar^2/(2md^2$). (In the case of a semiconductor billiard
referred to in the introduction, the mass $m$ should be the effective conduction band
mass $m^*$). Below we consider the specific case of small wave lengths $\lambda $ as
shown in Fig.~\ref{psi}. We will also comment on the case when $\lambda $ is large
compared to the dimensions of the cavity.

\begin{figure}[ht]
\includegraphics[width=.9\columnwidth]{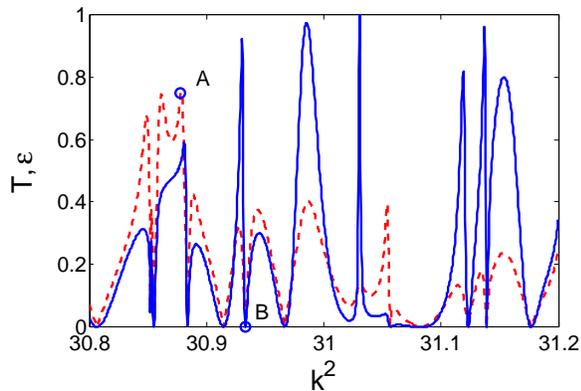}
\caption{ (color online)
The transmission probability $T$ (solid line) and $\epsilon$ (dashed line) as function
of the dimensionless energy $k^2$ for the Sinai billiard in Fig.~\ref{psi}. Two open
circles labeled show the case with maximal $\epsilon=0.75$ (A) and with the minimal
$\epsilon=0$ (B). At most only one channel is open in the leads.}
\label{transm}
\end{figure}

To ensure that the scattering wave function complies with a complex RGF we consider a
small aspect ratio $d/L$ as in Fig.~\ref{psi} (see also Ref.~[\onlinecite{ber02a}]).
The actual numerical size of the Sinai billiard in Fig.~\ref{psi} is chosen as:
height 346 (along transport), width ($L$) 670, radius 87, and 20 for the number of
grid points across the wave guides ($d$). Within this configuration we now only excite
 scattering wave function with characteristic wave lengths $\lambda \ll L $.
As expected from Fig.~\ref{psi} the wave function statistics show that both real and
imaginary parts, $u$ and $v$, obey Gaussian statistics to a high degree of accuracy.
Results for transmission $T$ and $\epsilon$ are are shown in Fig.~\ref{transm}.

\begin{figure}[ht]
\includegraphics[width=\columnwidth]{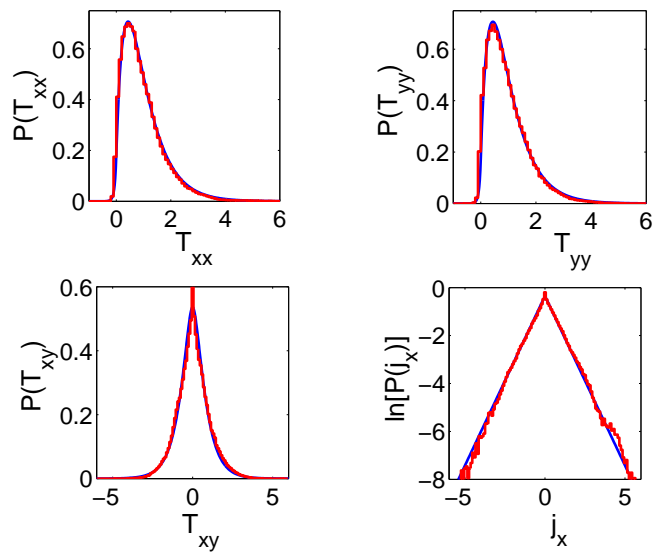}
\caption{ (color online)
Analytic and numerically simulated distributions of the components of the QST and probability
density current $j_x$ along the transport axis for the case A shown in Fig.~\ref{transm}
($\epsilon=0.75$). As in Figs.~\ref{fig1} and \ref{fig2} the diagonal components are measured
in terms of their mean values while $T_{xy}$ and $j_x$  are given in terms of
$\sqrt{\langle T_{xy}^2\rangle}$ and $\sqrt{\langle j_x^2\rangle}$, respectively. Solid lines
refer to analytic results for RGFs (Section~\ref{sectionIII} and Ref.~[\onlinecite{sai02b}])
and histograms to the present numerical modeling.  Because of the close agreement between
the two cases differences are barely resolved.}
\label{case-A}
\end{figure}

\begin{figure}[ht]
\includegraphics[width=\columnwidth]{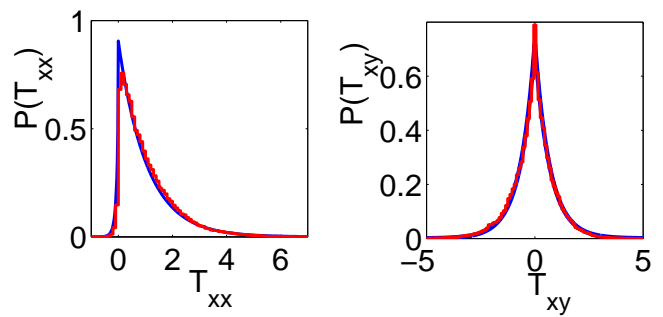}
\caption{ (color online)
Analytic and numerically simulated distributions of the components of the QST for the case B
in Fig.~\ref{transm} ($\epsilon=0$). The simulated distribution for $T_{yy}$ is nearly
identical to $P(T_{xx})$ and therefore not shown here. Because $\epsilon$ vanishes there
is not any current within the  cavity. (The choice of lines in the graphs and units  are
the same as in Fig.~\ref{case-A}. Because the close agreement between theory and simulations
differences are hardly noticeable.)}
\label{caseB}
\end{figure}

The corresponding distributions for the QST components are given in Figs.~\ref{case-A}
and \ref{caseB} supplemented by the distributions for  $j_x$ with the $x$ axis directed
along transport. There is indeed an overall good agreement between theory and simulations.
However, in the statistics for $j_x$ in Fig.~\ref{case-A} one notices a tiny difference at
small values of $j_x$. The reason is that there is a net current at this value of $\epsilon$,
which is not incorporated in our choice of analytic isotropic RGFs. The deviation is, however, much
too small to have an impact on the statistical analysis presented here because the net current
is such a tiny fraction of the entire current pattern within the cavity. The case B with
$\epsilon=0$ implies that the scattering wave function in the cavity is real (standing wave
with transmission $T=0$ as seen from Fig.~\ref{transm}). Therefore there is no current
within the cavity.

\begin{figure}[ht]
\includegraphics[width=\columnwidth]{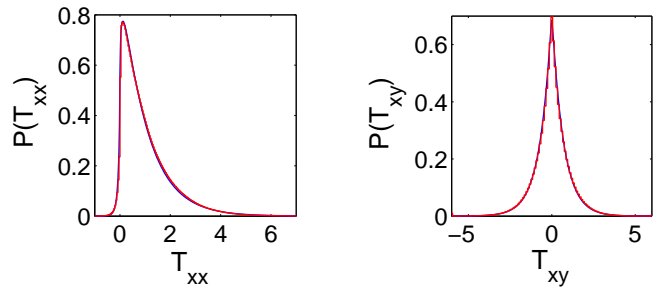}
\caption{ (color online)
Analytic and numerically simulated distributions of the components of stress tensor
$T_{xx}$ and $T_{xy}$ averaged over the energy window given in Fig.~\ref{transm}.
The theoretical curves are obtained also by averaging over computed
$\epsilon$-values
shown in Fig.~\ref{transm}. (The choice of lines and units are the same as in
Fig.~\ref{case-A}. The agreement between theory and simulations is, however,
excellent, hence any small differences are not resolved on the scale shown here.) }
\label{aver}
\end{figure}

The agreement with the analytic results for RGFs and the present numerical modeling for
billiards of finite size is obviously good in the range of energies explored here.
In order to smooth fluctuations in the distributions of the stress tensor we have
averaged over the energy window shown in Fig.~\ref{transm}
 (without scaling $\epsilon$ to 1 in contrast to
Fig.~\ref{fig::ExpSTDistr} of Section~\ref{exp}).
In this way one finds a perfect agreement
between theory and numerical simulations as shown in Fig.~\ref{aver}. For future
reference we note that the presence of net currents through the billiard appears to
have little or no influence on the distributions for the present two-lead configuration
and choice of energy range. We also note that the present results are not sensitive to
the position of the leads. For example, we have also performed simulations for Sinai
billiards with one dent only and with the leads attached at corners.

We now turn to the complementary case of long wave lengths (low energies). The low
energy regime is achieved for large aspect ratio $d/L$ which selects wave functions
with $\lambda$ a few times less than $L$. Moreover a low energy incoming wave often
excites bouncing modes. Numerics for the case $k^2=12$ and large aspect ratio
$d/L=1/7$ show that the scattering wave function may be rather different from a
complex RGF. Hence the corresponding distributions for individual states deviate
appreciably from the theoretical RGF predictions in Section~\ref{sectionIII}.
However, by averaging over a wide energy window, as above, one closes in on theory.
In this way one introduces an ensemble that, for practical purposes, mimics the
random Gaussian case. This aspect may be useful in experimental circumstances in
which the short wave length limit might be hard to achieve.

\section{Experimental studies}
\label{exp}

\begin{figure}[ht]
\includegraphics[width=.6\columnwidth]{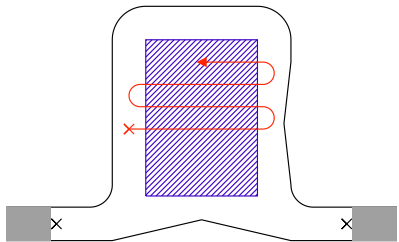}
\caption{(color online)
Sketch of the microwave billiard. The basic size of the billiard is
$\unit[16]{cm}\times\unit[21]{cm}$. The attached leads have a width of \unit[3]{cm}.
The central shaded field ($\unit[10]{cm}\times\unit[14]{cm}$) indicates the region
where the data have been collected. The measurement grid size was \unit[2.5]{mm}.
The gray regions at the end of the two leads indicate absorbers to mimic infinitely
long channels. The crosses indicate the antennas in the system and the winding path
illustrates how the third probing antenna is moved across the billiard during
measurements.}
\label{fig::ExpSetup}
\end{figure}

In quasi-2D resonators there is, as outlined in the introduction, a one-to-one
correspondence between the TM modes of the electromagnetic field and the wave
functions of the corresponding quantum billiard [\onlinecite{stoe99}]. The
$z$-component of the electromagnetic field $E_z$ corresponds to the
quantum-mechanical wave function $\psi$, and the wave number
$k^2 = \omega^2/c^2$ to the quantum-mechanical eigenenergy, where $\omega$ is
the angular frequency of the TM mode and $c$ the speed of light. In the present
study a rectangular cavity ($\unit[16]{cm}\times\unit[21]{cm}$) with rounded
corners has been used, with two attached leads with a width of \unit[3]{cm}.
Antennas placed in the leads  acted as source and drain for the microwaves
(see Fig.~\ref{fig::ExpSetup}). Two wedge-shaped obstacles had been attached
to two sides of the billiard to avoid any bouncing ball structures in the
measurement. The same system has been used already for the study of  a number
of transport studies [\onlinecite{bar02,kim03b}] and for the statistics of
nodal domains and vortex distributions [\onlinecite{kuh07b}]. A more detailed
description of the experimental setup can be found in Ref.~[\onlinecite{kuh00b}].
The field distribution inside the cavity has been obtained via a probe antenna
moved on a grid with a step size of \unit[2.5]{mm}. To avoid boundary effects,
only data from the shaded region (see Fig.~\ref{fig::ExpSetup}) has been
considered in the analysis.

\begin{figure}[t]
\includegraphics[width=.8\columnwidth]{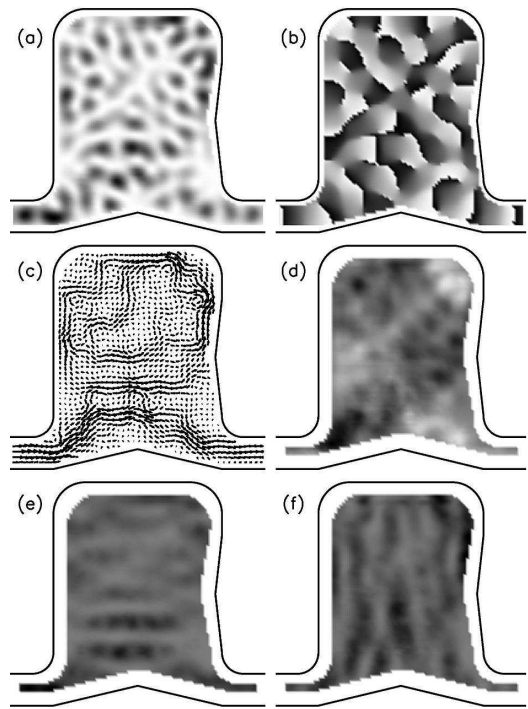}
\caption{
The figure shows different quantities obtained from the measurement at the
frequency $\nu=\unit[8.5]{GHz}$. In (a) the intensity of the wave function
is shown and in (b) its phase. The plot (c) shows the Poynting vector of the
system being equivalent to the probability current density in quantum
mechanics. In (d)-(f)  different components of the QST are shown, namely $xy$
(d),  $xx$ (e) and  $yy$ component (f). Dark areas indicate  higher values.}
\label{fig::Multiplot}
\end{figure}

The transmission from the source to the probe antenna has been measured on the
frequency range from 5.5 to \unit[10]{GHz} with a step size of \unit[20]{MHz},
corresponding to wave lengths from 3 to 5 cm. The transmission is proportional
to the electric field strength, i.\,e.\ to the wave function, at the position
of the probe antenna. This assumes that the leak current into the probe antenna
may be neglected.

To check this we compared the experimentally obtained distribution of wave
function intensities $\rho=\left|\psi\right|^2$ with the modified Porter-Thomas
distribution (see e.\,g.~Ref.~[\onlinecite{sai02b}])
\begin{equation}\label{pt}
 p\left(\left|\psi\right|^2\right) = \mu
 \exp\left(-\mu^2\left|\psi\right|^2\right)\,I_0\left(\mu\sqrt{\mu^2-1}\left|\psi\right|^2\right)
\end{equation}
where
\begin{equation}
 \mu=\frac{1}{2}\left(\epsilon+\frac{1}{\epsilon} \right)
 \qquad \textnormal{and}\qquad \epsilon ^2= \langle
 v^2\rangle/\langle u^2\rangle
\end{equation}
Here $\epsilon$ has not been fitted, but was taken directly from the experimentally
obtained values for $\langle u^2\rangle$ and $\langle v^2\rangle$, where we have
ensured that $\langle u v\rangle=0$ by applying a proper phase rotation as
in [\onlinecite{sai02b}] and commented on in Section~\ref{sectionV}). Whenever
$\chi^2$, the weighted squared difference of the experimental data and the modified
Porter-Thomas distribution, was below $\chi_{\mathrm{cutoff}}$=1.1, the pattern has
been selected for the final analysis of the statistics for the QST components.

Since the wave functions are experimentally known, including their phases, the
quantum-mechanical probability density ${\bf j}=\mathrm{Im} \psi^*{\bf\nabla} \psi$,
and the components of the QST can be obtained from the measurement. As mentioned,
distributions of current densities and related quantities have already been discussed
previously in a number of papers (see e.\,g.~Ref.~[\onlinecite{bar02,kim03b}]),
but the QST has not been studied experimentally before. As an example Fig.~\ref{fig::Multiplot}
show intensity (a) and the phase (b) of the measured field at one frequency, as well as
the probability current (c) and different components of the stress tensor (d - f).

The analysis of the data has been performed in dimensionless coordinates ${\bf x}=k{\bf r}$.
Since $u$ and $v$ are two independent random wave fields we may rescale the imaginary part
to obtain $\epsilon$ values of one, thus mapping the experimental result to the situation
of a completely open billiard. This step made it easy to superimpose the results from many
field patterns of different frequencies which originally had different $\epsilon$ values.
For the analysis all wave functions passing the $\chi^2$ test mentioned above have been used.
Altogether 83 of 225 possible patterns have been taken in the analysis.

\begin{figure}[t]
\includegraphics[width=\columnwidth]{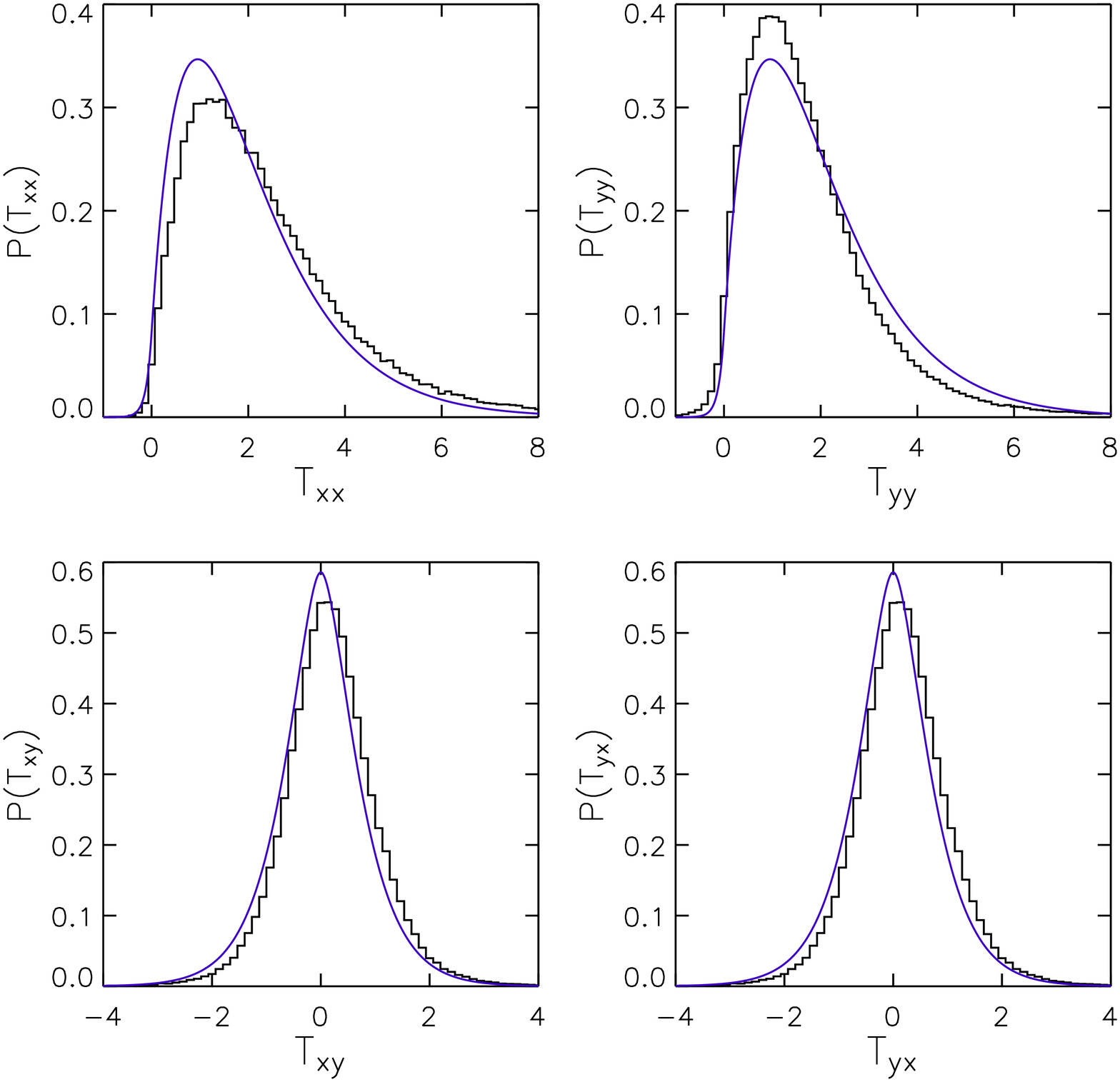}
\caption{ (color online)
Results for the experimental statistical distributions for the components of the QST stress
tensor obtained by a superposition of all experimental data scaled to $\epsilon =1$ as
explained in the text. The solid lines correspond to the theoretical predictions in
Section~\ref{sectionIII} for $\epsilon =1$.}
\label{fig::ExpSTDistr}
\end{figure}

Figure~\ref{fig::ExpSTDistr} shows the distribution of the QST components obtained in this way.
In addition the theoretical curves are shown as solid lines. From the figure we see that there
is a good overall agreement between experiment and theory, but also that non-statistical
deviations are unmistakable.

Deviations between experiment and theory had already been found by us  in the past in an
open microwave billiard, similar to the one used in the present experiment, in the
distribution of current components [\onlinecite{bar02,kim03b}]. For the vertical $y$
component a complete agreement between experiment and theory was found, but for the
horizontal $x$ component the experimental distribution showed, in contrast to theory,
a pronounced skewness. The origin of this discrepancy was a net current from the left
to the right due to source and drain in the attached wave guides. In a billiard with
broken time-reversal symmetry without open channels, a complete agreement between
experiment and theory had been found, corroborating the net current hypothesis.

\begin{figure}[t]
\includegraphics[width=.95\columnwidth]{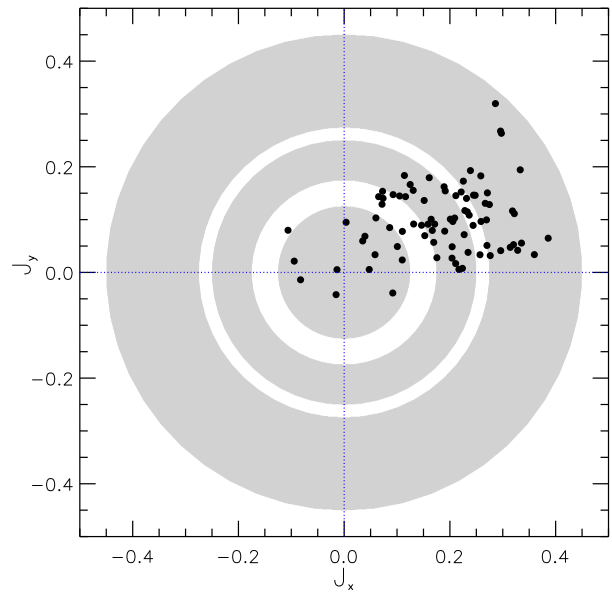}
\caption{
Plot of the net current as it is defined in Eq.~(\ref{eqn::DefNetCurrent}).
The shaded regions are indicating three different regimes of net current strength
which had been used in the later analysis.}
\label{fig::ScatterNetJ}
\end{figure}

For a quantitative discussion of the net current we introduced  the
normalized net current for each pattern
\begin{equation}
\label{eqn::DefNetCurrent} \mathbf{j}_{\mathrm{net}}=
\frac{\langle\mathbf{j}\rangle}{\langle|\mathbf{j}|\rangle}.
\label{directional}
\end{equation}
where the average is over all positions in the shaded region in
Fig.~\ref{fig::ExpSetup}. In Fig.~\ref{fig::ScatterNetJ} the $y$
component of $\mathbf{j}_{\mathrm{net}}$ is plotted versus its $x$
component for each wave function. One notices an average net current
pointing from left to right, with an angle of about twenty degrees
in an upward direction. For the analysis we discriminated between three
regimes for the strength of the net current. Additionally we
performed a coordinate transformation such that for each pattern the
vector of the net current is aligned along the positive $x$-axis.
This rotation has been done for all experimental and numerical
results in this section.

\begin{figure}[t]
\includegraphics[width=\columnwidth]{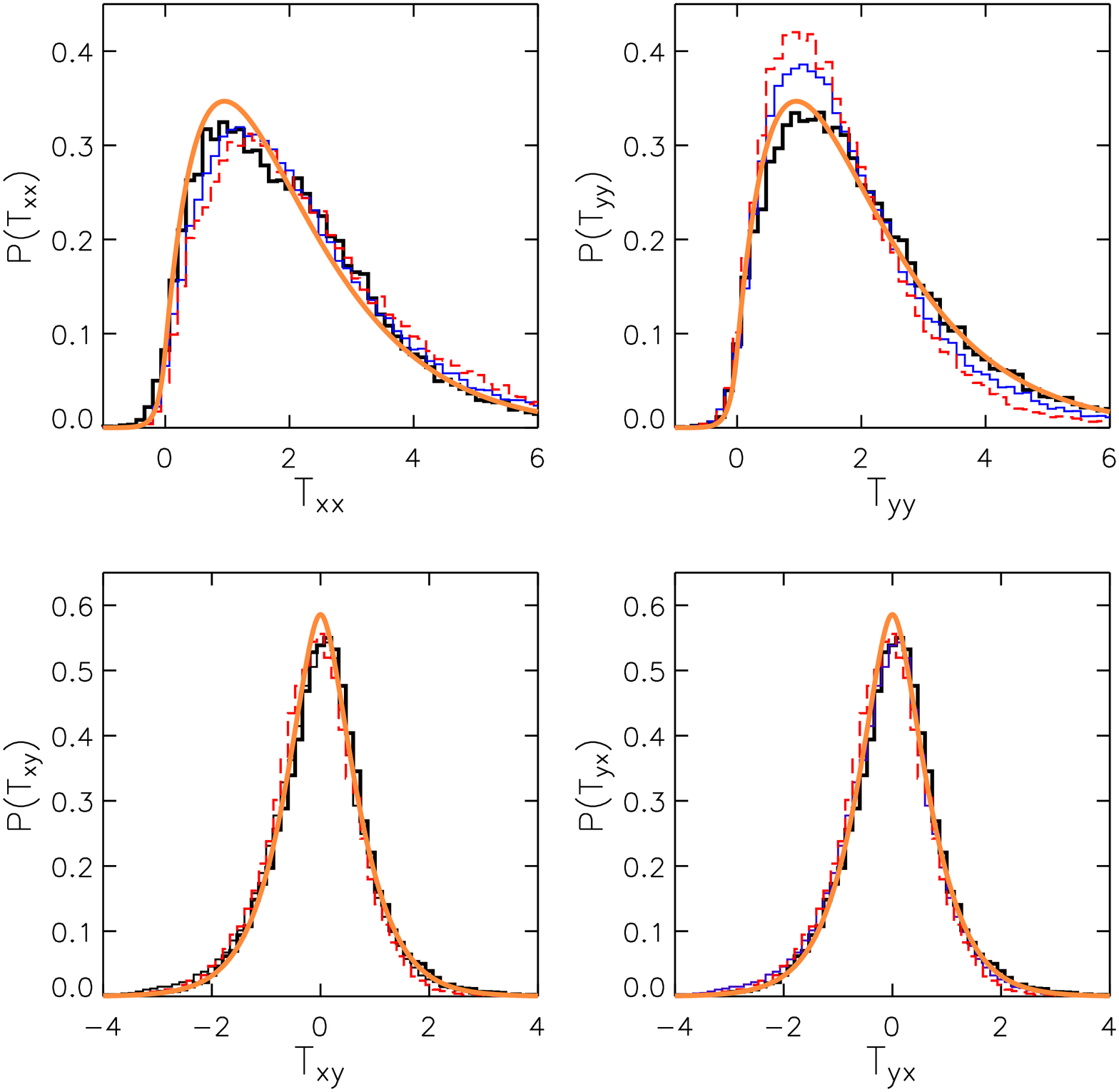}
\caption{ (color online)
Histograms of the QST distributions obtained from experimental data. The thick
lines corresponds to the smallest net currents (see Fig.~\ref{fig::ScatterNetJ}),
the thin lines to intermediate ones and the dashed lines to ones with the largest
net current. As in Fig.~\ref{fig::ExpSTDistr} the solid lines correspond to the
theoretical predictions in Section~\ref{sectionIII} for $\epsilon =1$. }
\label{fig::ExpSTDistrAll}
\end{figure}

In Fig.~\ref{fig::ExpSTDistrAll} the results for the three different regimes of
net current strengths are shown. For the distributions of the $xx$ and the $yy$
component of the QST, a clear dependence on the net current strength is found,
where the deviations from theory increase with an increasing net current. $T_{xy}$ is
only slightly affected by the net current, if at all. In the limit of a tiny net
current, all experimental distributions approach the theoretical ones.

\begin{figure}[t]
\includegraphics[width=\columnwidth]{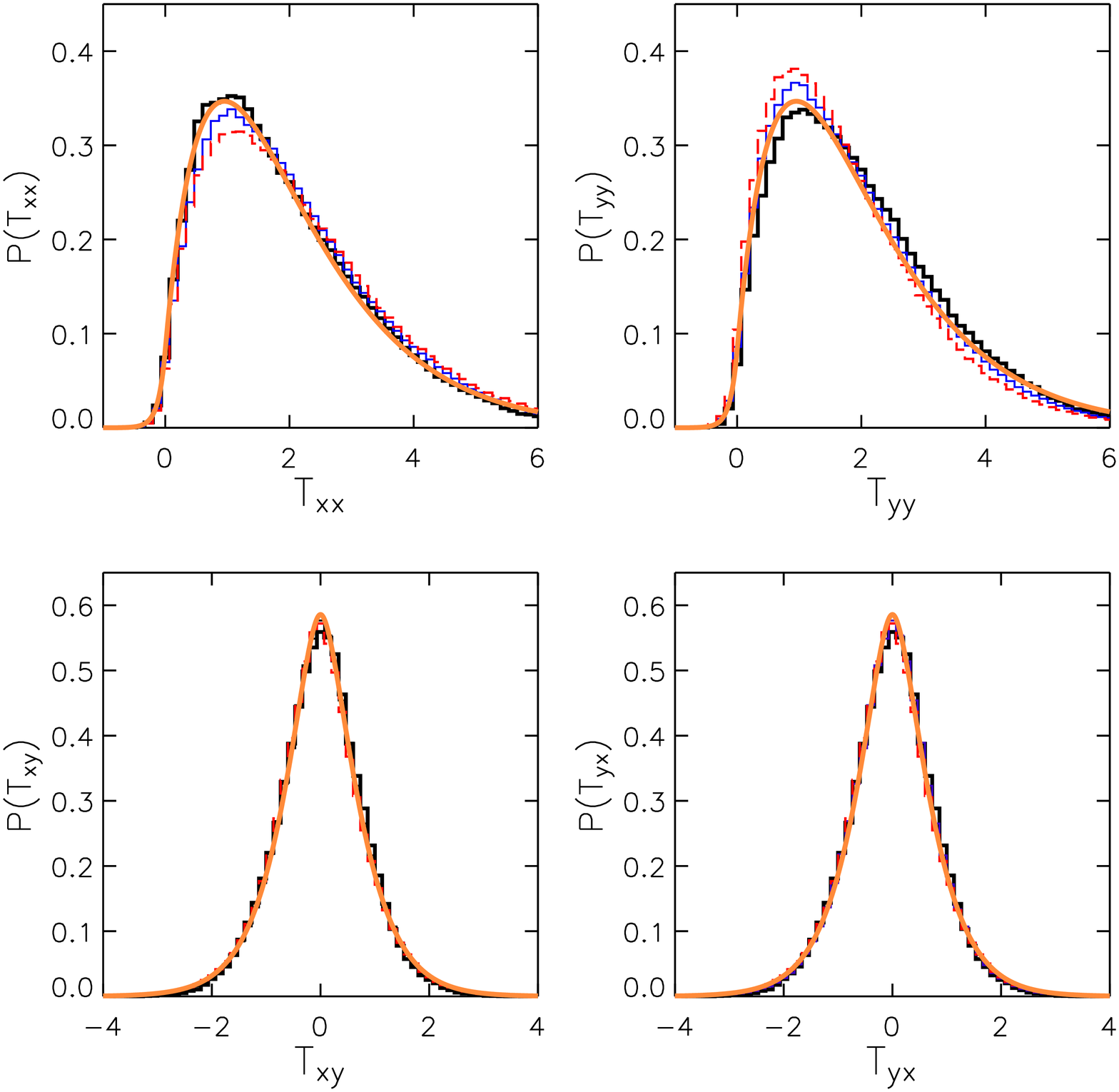}
\caption{ (color online)
Histograms of the QST components obtained from the simulations according to the wave
function in Eq.~(\ref{eqn::NumRSPW}). As in previous figure, the thick lines correspond to low, thin
to intermediate and dashed lines to large net currents.}
\label{fig::RSPWRotatedDistr}
\end{figure}

To further test  the influence of the net current on the distributions of the stress
tensor, we performed a numerical simulation with random plane waves. Each random wave
field was calculated on an area of $\unit[500]{mm} \times \unit[500]{mm}$, with a
grid size of \unit[2.5]{mm}. The random wave field consisted of 500 plane waves with
random directions and amplitudes. The frequency used for the numerics was
$\nu = \unit[5]{GHz}$. To introduce the net current we first performed a random
superposition of plane waves according to Eq.~(\ref{Berry}), and then added a
normalized plane wave with the wave vector $K'$ pointing in the same direction as the
net current observed in the experiment,
\begin{equation}
 \psi(r)=\frac{1}{\sqrt{A}}\left(a'e^{i{\bf K'\cdot r}} + \sum_{n=1}^N a_n e^{i{\bf k}_n{\bf \cdot r}}\right)
 \label{eqn::NumRSPW}
\end{equation}
The strength of the resulting net current was adjusted by a prefactor $a'$. The best
agreement between the experiment and the numerics was found for $a'=0.45$. To get
sufficient statistics we averaged over 200 different wave functions.  Thus a pattern similar to the one shown in Fig. \ref{fig::ScatterNetJ} was obtained with a cloud of dots extending over all three regimes of net current considered with its center in the central regime.

Fig.~\ref{fig::RSPWRotatedDistr} shows the distributions for the QST components for
numerical data derived from Eq.~(\ref{eqn::NumRSPW}). The same three regimes as for
the experimental study have been used. The results from this type of simulation are
in good qualitative agreement with the experimental results. In particular the
deviations from the theory in Section~\ref{sectionIII} increase monotonously with the
net current, just as in the experiment.

An obvious question is why these net current effects are unimportant in the simulations
for the Sinai billiard presented in Section~\ref{sectionV}. One may argue that the
number of independent plane waves entering at a given frequency is given by the
circumference of the billiard divided by $\lambda/2$, where $\lambda$ is the wavelength.
Also the width of each wave guide is of the order of $\lambda/2$, i.\,e. the relative net
current is approximately  given by  the total widths of all openings divided by the
circumference of the billiard. Following this argumentation  the net current in the
experiments amounted to about 10~percent of the total current, whereas in the simulations
for the Sinai billiard it was smaller by a factor of 10; i.e. too small to be of any importance
in the simulations.

We have shown that  in the limit of  small net currents, the distributions of QST components
obtained from the experiment  are well described by means of the random plane wave model
and the analytic distributions in Section~\ref{sectionIII}. On the other hand net currents
are unavoidable in open systems. As indicated by the simulations for a Sinai billiard in
Section~\ref{sectionV}, the magnitudes and effect on the different stress tensor
distributions may be sensitive to geometry and energy. Hence it remains an open task for
theory to incorporate net currents in order to allow a more realistic comparison with present
experimental results.

\section{Acknowledgements}
\label{ackn}

The theoretical part of this joint project has been carried out by the two theory groups at
Link\"{o}ping University and Kirensky Institute of Physics. Measurements and the data interpretation
have been performed by the quantum chaos group at the Philipps-Universit\"at Marburg.

K.-F.~B., D.~M. and A.F.~S. are grateful to the Royal Swedish Academy of Sciences for
financial support for the theory part of this work (``Academy Programme for Collaboration
between Sweden and Russia''). R.~H., U.~K. and H.-J.~S.\ thank the Deutsche
Forschungsgemeinschaft for financial support of the experiments (via Forschergruppe 760
``Scattering systems with complex dynamics"). Finally, KFB is grateful to Andrew W. Rappe,
Jianmin Tao and Irina I. Yakimenko for informative discussions on the concept of quantum stress.

\end{document}